# Magnetic Fingerprints of sub-100 nm Fe Nanodots

Randy K. Dumas,[1] Chang-Peng Li,[2] Igor V. Roshchin,[2] Ivan K. Schuller[2] and Kai Liu[1,*]

[1]*Physics Department, University of California, Davis, California 95616*

[2]*Physics Department, University of California - San Diego, La Jolla, California 92093*

## Abstract

Sub-100 nm nanomagnets not only are technologically important, but also exhibit complex magnetization reversal behaviors as their dimensions are comparable to typical magnetic domain wall widths. Here we capture magnetic "fingerprints" of $10^9$ Fe nanodots as they undergo a single domain to vortex state transition, using a first-order reversal curve (FORC) method. As the nanodot size increases from 52 nm to 67 nm, the FORC diagrams reveal striking differences, despite only subtle changes in their major hysteresis loops. The 52 nm nanodots exhibit single domain behavior and the coercivity distribution extracted from the FORC distribution agrees well with a calculation based on the measured nanodot size distribution. The 58 and 67 nm nanodots exhibit vortex states, where the nucleation and annihilation of the vortices are manifested as butterfly-like features in the FORC distribution and confirmed by micromagnetic simulations. Furthermore, the FORC method gives quantitative measures of the magnetic phase fractions, and vortex nucleation and annihilation fields.

**PACS's: 75.60.Jk, 75.60.-d**, 75.70.Kw, 75.75.+a

## I. Introduction

Deep sub-100 nm magnetic nanoelements have been the focus of intense research interest due to their fascinating fundamental properties and potential technological applications.[1-6] At such small dimensions, comparable to the typical magnetic domain wall width, properties of the nanomagnets are rich and complex. It is known that well above the domain wall width, in micron and sub-micron sized patterns, magnetization reversal often occur via a vortex state (VS).[7-12] At reduced sizes, single domain (SD) static states are energetically more favorable.[13] However, even in SD nanoparticles, the magnetization reversal can be quite complex, involving thermally activated incoherent processes.[14] The VS-SD crossover itself is fascinating. For example, recently Jausovec *et al.* have proposed that a third, metastable, state exists in 97 nm permalloy nanodots, based on minor loop and remanence curve studies.[15] To date, direct observation of the VS-SD crossover, especially in the deep sub-100 nm regime, has been challenging. Fundamentally, the vortex core is expected to have a nanoscale size, comparable to the exchange length. Magnetic imaging techniques face resolution limits and often are limited to remanent state and room temperature studies. Practically, collections of nanomagnets inevitably have variations in size, shape, anisotropy, etc.[16] The ensemble-averaged properties obtained by collective measurements such as magnetometry no longer yield clear signatures of the nucleation / annihilation fields. Furthermore, quantitatively capturing the distributions of magnetic properties are essential to the understanding and application of magnetic and spintronic devices, which may consist of billions of nanomagnets. How to *qualitatively and quantitatively* investigate the properties of such nanomagnets remains a key challenge for condensed matter physics and materials science.

In this study, we investigate the VS-SD crossover in deep sub-100 nm Fe nanodots. We have captured "fingerprints" of such nanodots using a first-order reversal curve (FORC) method,[17-21] which circumvents the resolution, remanent state and room temperature limits by measuring the collective magnetic responses of the dots. The "fingerprints", shown as FORC diagrams, reveal remarkably rich information about the nanodots. A qualitatively different reversal pattern is observed as the dot size is increased from 52 to 67 nm, despite only subtle differences in their major hysteresis loops. The 52 nm nanodots behave as SD particles; the 67 nm ones exhibit VS reversal; and the 58 nm ones have both SD and VS characteristics. Quantitatively, the FORC diagram shows explicitly a coercivity distribution for the SD dots, which agrees well with calculations; it yields SD and VS phase fractions in the larger dots; it also extracts unambiguously the nucleation and annihilation fields for the VS dots and distinguishes annihilations from opposite sides of the dots.

**II. Experimental**

Samples for the study are Fe nanodots fabricated using a nanoporous alumina shadow mask technique in conjunction with electron beam evaporation.[22,23] This method allows for fabrication of high density nanodots (~$10^{10}$ /cm$^2$) over macroscopic areas (~1 cm$^2$). Three different types of samples have been made on Si and MgO substrates with mean nanodot sizes of 52±8, 58±8, and 67±13 nm, and a thickness of 20 nm, 15 nm, and 20 nm, respectively. The nanodot center-to-center spacing is typically twice its diameter.

The Fe nanodots thus made are polycrystalline, capped with an Al or Ag layer. A scanning electron microscopy (SEM) image of the 67 nm sample is shown in Fig. 1. A survey of the size distribution is illustrated in Fig. 1 inset.

Magnetic properties have been measured using a Princeton Measurements Corp. 2900 alternating gradient and vibrating sample magnetometer (AGM/VSM), with the applied field in the plane of the nanodots. Samples have been cut down to ~ 3 × 3 mm$^2$ pieces, which contains ~ 10$^9$ Fe nanodots each. Additionally, the FORC technique has been employed to study details of the magnetization reversal. After saturation, the magnetization $M$ is measured starting from a reversal field $H_R$ back to positive saturation, tracing out a FORC. A family of FORC's is measured at different $H_R$, with equal field spacing, thus filling the interior of the major hysteresis loop [Figs. 2(a)-2(c)]. The FORC distribution is defined as a mixed second order derivative:[17-21]

$$\rho(H_R, H) \equiv -\frac{1}{2} \frac{\partial^2 M(H_R, H)}{\partial H_R \partial H}, \qquad (1)$$

which eliminates the purely reversible components of the magnetization. Thus any non-zero $\rho$ corresponds to *irreversible* switching processes.[19-21] The FORC distribution is plotted against ($H$, $H_R$) coordinates on a contour map or a 3-dimensional plot. For example, along each FORC in Fig 4(a) with a specific reversal field $H_R$, the magnetization $M$ is measured with increasing applied field $H$; the corresponding FORC distribution $\rho$ in Fig. 4(b) is represented by a horizontal line scan at that $H_R$ along $H$. Alternatively $\rho$ can be plotted in coordinates of ($H_C$, $H_B$), where $H_C$ is the local coercive field and $H_B$ is the local interaction or bias field. This transformation is accomplished by a simple rotation of the coordinate system defined by: $H_B=(H+H_R)/2$ and $H_C=(H-H_R)/2$. Both coordinate systems are discussed in this paper.

**III. Results**

Families of the FORC's for the 52, 58, and 67 nm nanodots are shown in Figs. 2(a)-2(c). The major hysteresis loops, delineated by the outer boundaries of the FORC's, exhibit only subtle differences. The 52 nm nanodots show a regular major loop, with a remanence of 57 % and a coercivity of 475 Oe [Fig. 2(a)]. The 67 nm nanodots have a slight "pinching" in its loop near zero applied field, with a remanence of 27 % and a coercivity of 246 Oe [Fig. 2(c)]. The unique shape, small values of coercivity and remanence suggest that the magnetization reversal is via a VS. Indeed, the VS is confirmed by polarized neutron reflectivity measurements on similarly prepared 65 nm Fe nanodots, which find an out of plane magnetic moment corresponding to a vortex core of 15 nm.[24] However, due to the relatively gradual changes in magnetization along the major loop, averaged over signals from ~ $10^9$ Fe nanodots, it is difficult to determine the vortex nucleation and annihilation fields. In contrast, the relatively fuller major loop of the 52 nm nanodots is suggestive of a SD state. The loop of the 58 nm nanodots appears to have combined features from those of the other two samples [Fig. 2(b)].

The subtle differences seen in the major hysteresis loops manifest themselves as striking differences in the corresponding FORC distributions, shown in Figs. 2(d)-2(i). For the 52 nm nanodots, the only predominant feature is a narrow ridge along the local coercivity $H_C$-axis with zero bias [Figs. 2(d) and 2(g)]. The ridge is peaked at $H_C$ = 525 Oe, near the major loop coercivity value of 475 Oe. This pattern is characteristic of a collection of non-interacting SD particles.[25] Given that the nanodot spacing is about twice its diameter and the random in-plane easy axes, dipolar interactions are expected to be small.[26] The relative spread of the FORC distribution along the $H_B$-axis actually gives a direct measure of the interdot interactions, as we have shown in single domain magnetite nanoparticles with different separations.[27] The sharp ridge shown in Fig. 2(g) is similar to that of an assembly of well–dispersed magnetite nanoparticles with little dipolar interactions. In the present case, the ridge is localized between bias field of $H_B \sim \pm 100$ Oe and has a narrow FWHM (full width at half maximum) of about 136 Oe [Fig. 3(a)].  A simple calculation of the dipolar fields yields a value similar to the FWHM.

As the nanodot size is increased, the FORC distribution becomes much more complex.  The 67 nm sample is characterized by three main features, as shown in Figs. 2(f) and 2(i): two pronounced peaks at $H_C = 650$ Oe and $H_B = \pm 750$ Oe, and a ridge along $H_B = 0$, forming a butterfly-like contour plot [Fig. 2(i)]. The ridge has changed significantly from that of the 52 nm sample: a peak corresponding to the coercivity of the major loop is now virtually absent; instead a large peak at $H_C = 1500$ Oe has appeared, accompanied by two small negative regions nearby.  The 58 nm sample shows a FORC pattern representative of both the 52 and 67 nm samples [Figs. 2(e) and 2(h)].  The overall distribution resembles that of the 67 nm sample, with two peaks centered at $H_C = 650$ Oe and $H_B = \pm 400$ Oe.  A ridge along $H_B = 0$, peaking at roughly the major loop coercivity, is similar to that seen in the 52 nm sample.  Note that the 58 nm sample, being thinner, would tend to inhibit the formation of a VS in the smaller dots and therefore show magnetic reversal via a SD.  However, significant fractions of the nanodots in the ensemble are apparently reversing via a VS.

The FORC distribution also allows us to extract quantitative information about the reversal processes.  Since each sample measured consists of ~ 1 billion nanodots with a distribution of sizes, a coercivity spread is contained in the FORC distribution.  For the 52 nm sample, we have indeed extracted this distribution by projecting the ridge in Fig. 2(g) onto the $H_C$-axis ($H_B = 0$), as shown as the open circles in Fig. 3(b). The relative height gives the appropriate weight of nanodots with a given coercivity. This extracted coercivity distribution can be compared with a simple theoretical calculation. The coercivity of a SD particle undergoing reversal via a curling mode increases strongly with decreasing particle size $d$, according to

$$H_C \propto \frac{C_1}{d^2} - C_2, \tag{2}$$

where $C_1$ and $C_2$ are constants.[28] Based on the mean nanodot size and the size distribution determined from SEM, we have calculated a coercivity distribution [solid circles in Fig. 3(b)]. A good agreement is obtained with that determined from the FORC distribution, after a rescaling of the latter by an arbitrary weight. Thus for other non-interacting single-domain particle systems with unknown size distributions, the FORC method may be used to extract that information. This is particularly important in 3D distributions of nanostructures where there is no direct image access to the individual dots, as is the case here for a 2D distribution.

As we have demonstrated earlier, the FORC distribution $\rho$ is extremely sensitive to irreversible switching.[19] This is most convenient to see in the ($H$, $H_R$) coordinate system (meaningful data is in $H>H_R$), as non-zero values of $\rho$ correspond to the degree of irreversibility along a given FORC. We have employed this capability to analyze the VS nucleation and annihilation for the 67 nm sample. The complex butterfly-like pattern of Fig. 2(i) now transforms into irreversible switching mainly along line scans **1** and **2** in Fig. 4(b), which correspond to FORC's starting at $H_R$= 100 Oe and -1450 Oe, respectively [marked as bold with large open circles as starting points in Fig. 4(a)]. Along line scan **1** ($H_R$=100 Oe), when applied field $H$=100 Oe, vortices have already nucleated in most of the nanodots. With increasing field $H$, $\rho$ becomes non-zero and increases with $H$ and peaks at 1320 Oe. This corresponds to the annihilation of the vortices in majority of the nanodots, and eventually $\rho$ returns to zero near positive saturation. Line scan **2** starts at $H_R$= -1450 Oe, where the majority, but not all, of the nanodots have been negatively saturated. As $H$ is increased, a first maximum in $\rho$ is seen at $H$= -100 Oe, corresponding to the nucleation of vortices within the nanodots. Between -100 Oe $< H <$ 1450 Oe, $\rho$ is essentially zero, indicating reversible motion of the vortices through the nanodots. A second $\rho$ maximum is found at $H$ = 1450 Oe, as the vortices are annihilated. This is again followed by reversible behavior near positive saturation. Note that along line scan **1**, the vortices are annihilated from the same side of the nanodot from which they first nucleated, and thus the net magnetization remains positive; along line scan **2**, the vortices nucleate on one side of the dot and are annihilated from the other, and consequently the net magnetization changes sign [Fig. 4(a)]. Interestingly the annihilation field along line scan **2**, 1450 Oe, is larger than that along line scan **1**, 1320 Oe. It seems more difficult to drive a vortex across the nanodot and then annihilate it. The peaks in Fig. 4(b) are rather broad, which is a manifestation of vortex nucleation and annihilation field distributions. Also note that the interactions among the VS dots are expected to be negligible due to the high degree of flux closure, as confirmed by simulations.[29]

**IV. Simulations and Discussions**

For comparison, micromagnetic simulations have been carried out on nearly circular nanodots with 60 nm diameters and 20 nm thicknesses.[30] We have used parameters appropriate for Fe (exchange stiffness A = 2.1 × $10^{-11}$ J/m, saturation magnetization $M_s$ = 1.7 × $10^6$ A/m, and anisotropy constant K = 4.8 × $10^4$ J/m$^3$). Each polycrystalline nanodot is composed of 2 nm square cells that are 20 nm thick, where each cell is a different grain with a random easy axis. A small cut on one side of the nanodot generates two distinct annihilation fields that depend on which side of the nanodot the vortex annihilates from. This exercise models the fact that our fabricated dots are not perfectly circular.[31] We have simulated FORC's generated by two nanodots with different orientations [Fig. 4(c)]: the edge-cut in one is parallel, and in the other at a 45º angle, to the applied field. The simulated *M-H* curves show abrupt magnetization changes, corresponding to the nucleation, propagation, and annihilation of vortices.

The corresponding FORC distribution is shown in Fig. 4(d). Peaks in the simulated FORC distribution clearly indicate the nucleation and annihilation fields of the vortices which are apparent in Fig 4(c). Along line scan **1** of Fig. 4(d), a vortex is already nucleated at $H_R$ = 100 Oe and subsequently annihilated at *H* = 2300 Oe (upper right corner). Along line scan **2** with $H_R$ = -2450 Oe, a vortex is nucleated at *H* = -100 Oe and finally annihilated at *H* = 2450 Oe (lower right corner). It is clear that the simulated FORC reproduces the key features of the experimentally obtained one in Fig. 4(b). Here the asymmetric dot shape is essential to obtain a different annihilation field along scan **2** than that along scan **1**. We have simulated the angular dependence of nucleation and annihilation fields in such circular dot with a small cut as the cut orientation is varied in a field. We find that for most angles it is harder to annihilate a vortex from the opposite side of its nucleation site. However, for a small range of angles near 45° it is actually slightly easier to annihilate from the opposite side. It is the combination of these two behaviors that gives rise to the negative-positive-negative trio of features in the lower right portion of the FORC distribution. The presence of similar features in the experimental data shown in Fig. 4(b) thus illustrates that the FORC distribution is also sensitive to variations of dot shapes in the array. Because only two dots are simulated, the features in the FORC distribution are much sharper than the experimental data where distributions of vortex nucleation and annihilation fields are present. Including more dots in the simulation with different applied field orientations and size distributions would tend to broaden the features generated by the two dots simulated.

Additionally, by selectively integrating the normalized FORC distribution[20,21] corresponding to the SD phase (the aforementioned ridge centered at low coercivity values in Fig. 2), we can quantitatively determine the percentage of nanodots in SD state for each sample. The SD phase fraction is 100%, 43%, and 10% for the 52, 58, and 67 nm sample, respectively. Thus the 58 nm nanodots have a significant co-existence of both SD and VS states. However, we do not observe clear evidence of any additional metastable phase.[15]

## V. Conclusions

In summary, we have used the FORC method to "fingerprint" the rich magnetization reversal behavior in arrays of 52, 58, and 67 nm sized Fe nanodots.

Distinctly different reversal mechanisms have been captured, despite only subtle differences in the major hysteresis loops. The 52 nm nanodots are in SD states. A coercivity distribution has been extracted, which agrees with calculations. The 67 nm dots reverse their magnetization via the nucleation and annihilation of vortices. Different fields are required to annihilate vortices from opposite sides of the dots. Quantitative measures of the vortex nucleation and annihilation fields have been obtained. OOMMF simulations confirm the experimental FORC distributions. The 58 nm sample shows coexistence of SD and VS reversal, without evidence of additional reversal mode. These results further demonstrate the FORC method as a simple yet powerful technique for studying magnetization reversal, due to its capability of capturing distributions of magnetic properties, sensitivity to irreversible switching, and the quantitative phase information it can extract.

**Acknowledgements**

This work has been supported by ACS (PRF-43637-AC10), AFOSR, and the Alfred P. Sloan Foundation. We thank J. E. Davies, J. Olamit, M. Winklhofer, C. R. Pike, H. G. Katzgraber, R. T. Scalettar, G. T. Zimányi, and K. L. Verosub for helpful discussions. R.K.D. acknowledges support from the Katherine Fadley Pusateri Memorial Travel Award.

**Figure Captions**

Fig. 1. (color online). Scanning electron micrograph of the 67 nm diameter nanodot sample. Inset is a histogram showing the distribution of nanodot sizes.

Fig. 2. First-order reversal curves and the corresponding distributions. Families of FORC's (a-c), whose starting points are represented by black dots for the 52, 58, and 67 nm Fe nanodots, respectively. The corresponding FORC distributions are shown in 3-dimensional plots (d-f) and contour plots (g-i).

Fig. 3. (color online). Projection of the FORC distribution $\rho$ of the 52 nm nanodots onto (a) the $H_B$-axis, showing weak dipolar interactions; and (b) the $H_C$-axis (open circles), showing a coercivity distribution that agrees with a calculation based on measured size distribution (solid circles).

Fig. 4. (a) A family of measured FORC's for the 67nm diameter dots. (b) The corresponding experimental FORC distribution plotted against applied field $H$ and reversal field $H_R$. (c) A family of simulated FORC's generated using the OOMMF code. Inset shows the orientations of the two dots simulated. (d) The FORC distribution calculated from the simulated FORC's shown in (c). The two white dashed lines in (b) and (d) correspond to the two bold FORC's whose starting points are large open circles in (a) and (c), respectively.

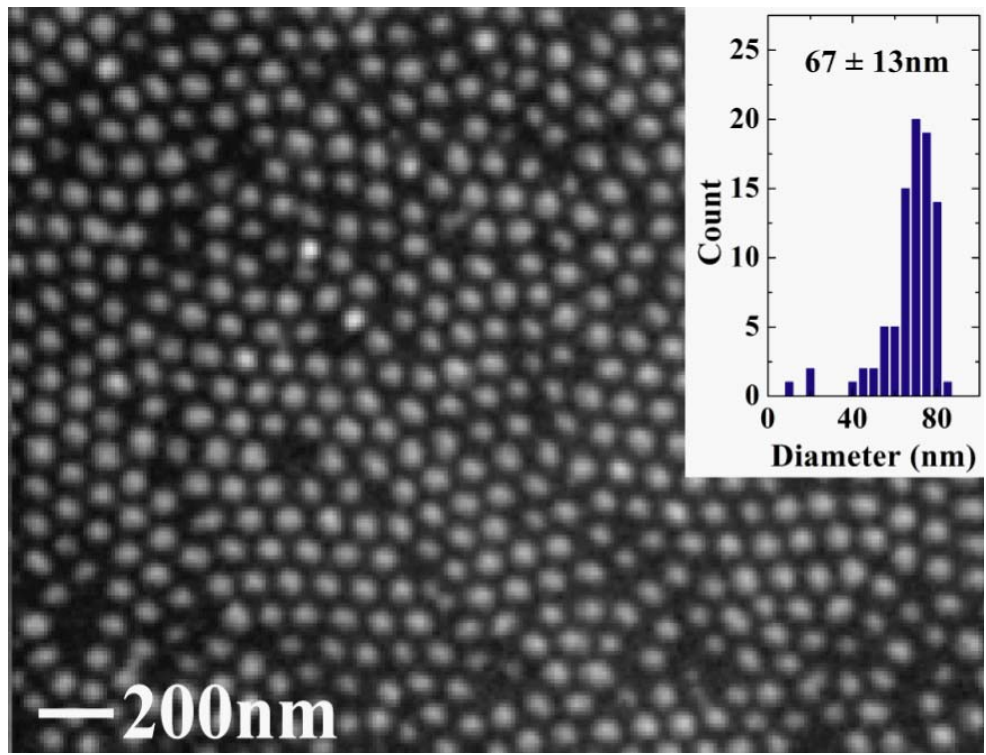

**Fig. 1, Dumas, et al.**

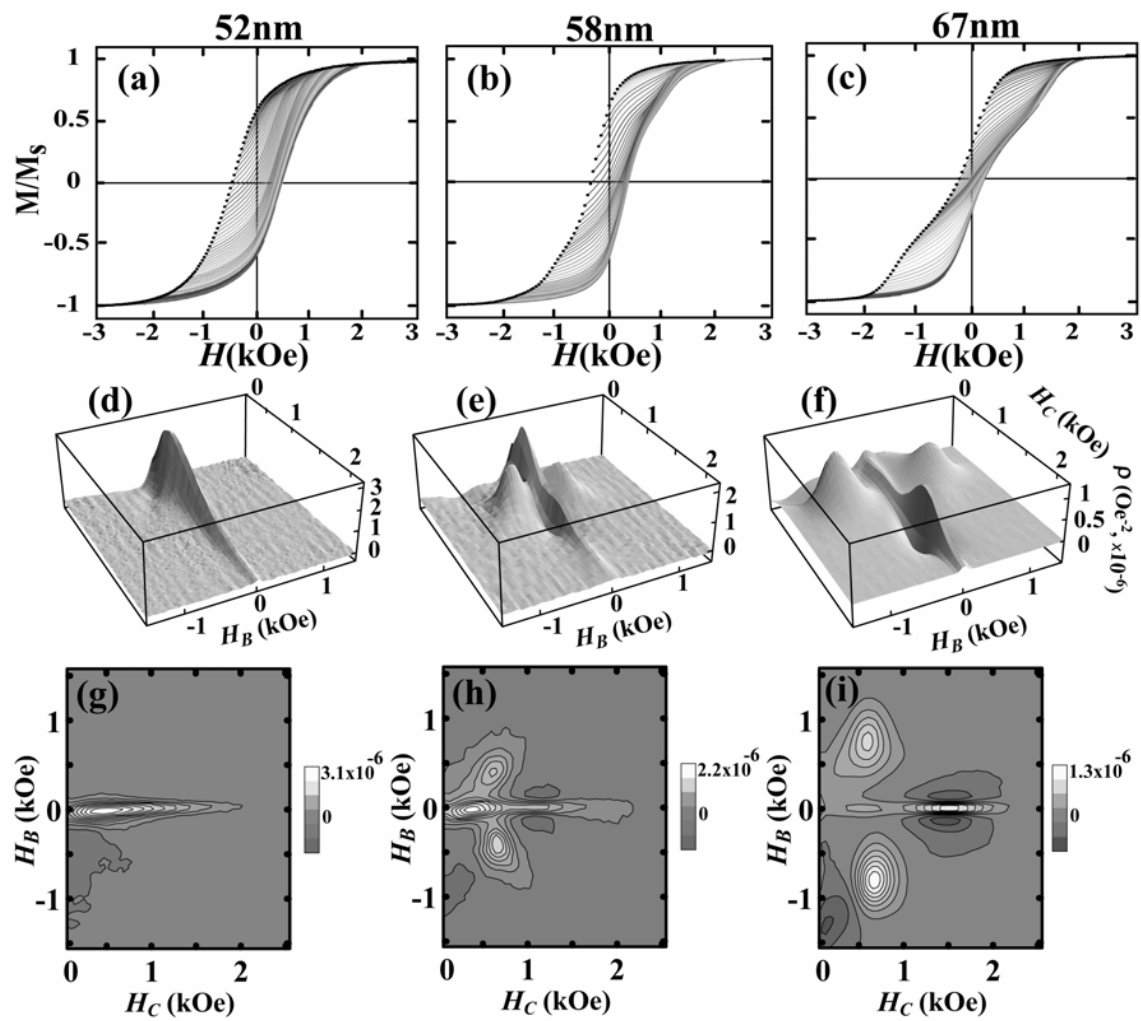

**Fig. 2, Dumas, et al.**

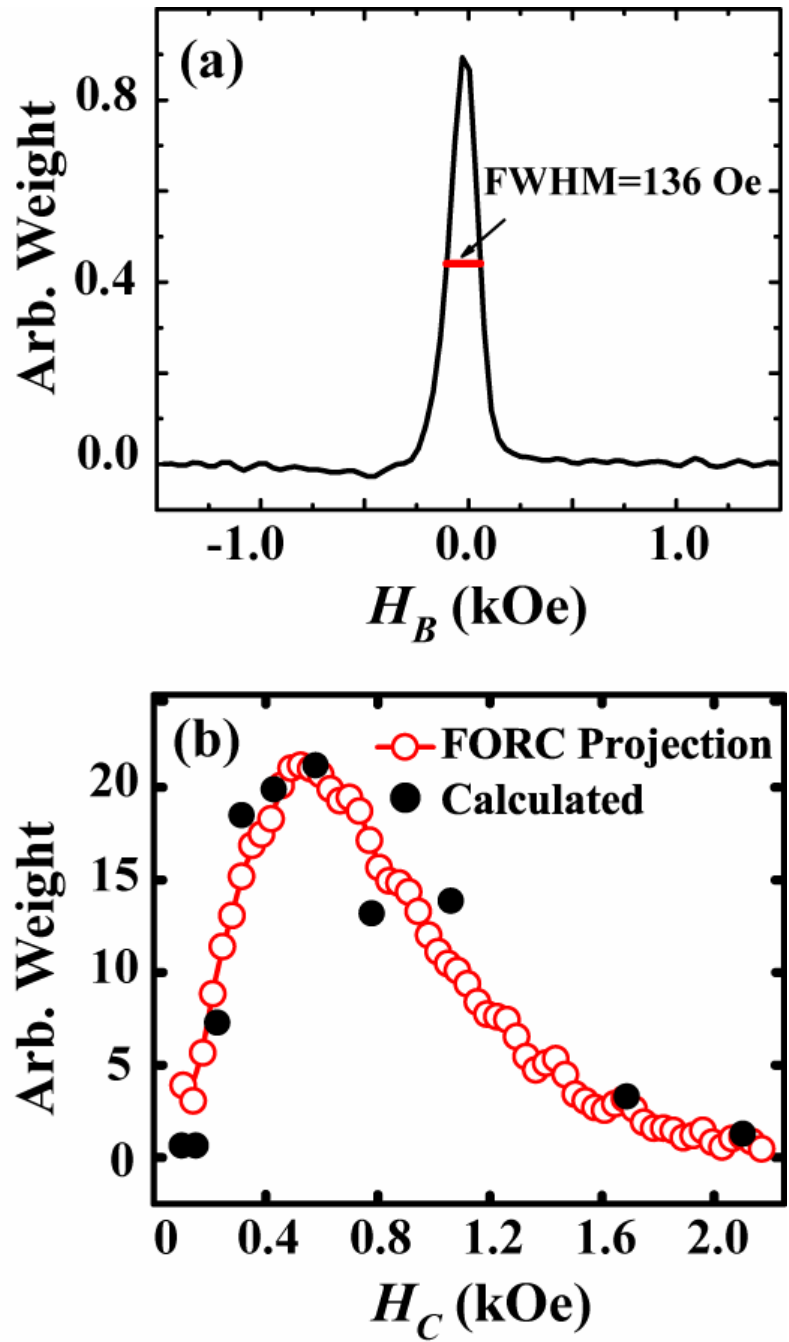

**Fig. 3, Dumas, et al.**

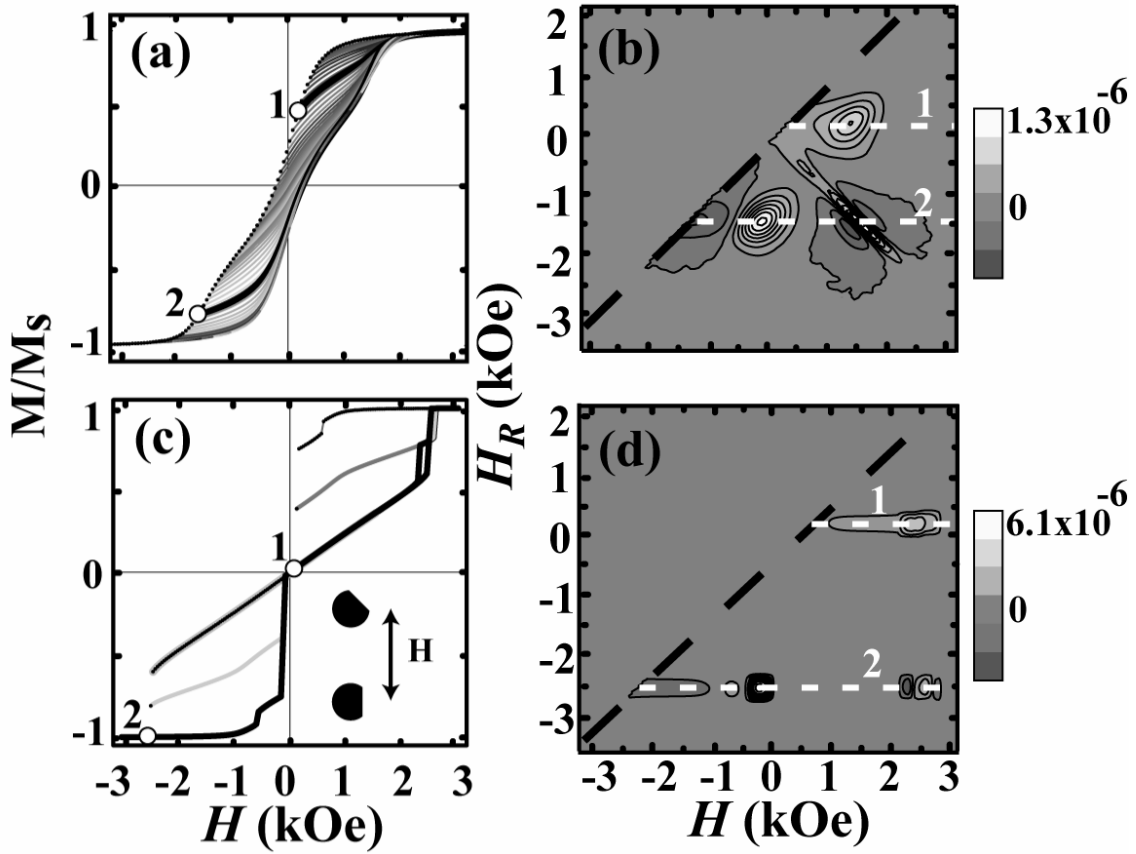

**Fig. 4, Dumas, et al.**

19